\documentclass[aps,prd,epsf,showpacs,amsmath,amssymb,graphics,
twocolumn,10pt]{revtex4}
\usepackage{graphicx}
\usepackage{epsfig}
\usepackage{dcolumn}
\usepackage{bm}

\newcommand{\be}{\begin{equation}}
\newcommand{\ee}{\end{equation}}
\newcommand{\ba}{\begin{eqnarray}}
\newcommand{\ea}{\end{eqnarray}}
\newcommand{\ban}{\begin{eqnarray*}}
\newcommand{\ean}{\end{eqnarray*}}
\newcommand{\bfg}{\begin{figure}}
\newcommand{\efg}{\end{figure}}

\newcommand{\subse}{\subsection}
\newcommand{\ct}{\cite}

\newcommand{\eq}[1]{(\ref{#1})}

\def\th{\theta}

\def\lt{\left}
\def\rt{\right}

\def\fr{\frac}

\newcommand{\gr}{General Relativity     }

\newcommand{\sch}{Schwarzschild   }

\newcommand{\grle}{gravitational lensing     }

\newcommand{\mas}{microarcseconds     }

\newcommand{\eda}{Einstein deflection angle     }

\begin{document}

\title{Can strong gravitational lensing distinguish naked singularities from black holes?}

\author{Satyabrata Sahu\footnote{Electronic address: satyabrata@tifr.res.in},
Mandar Patil\footnote{Electronic address: mandarp@tifr.res.in},
D. Narasimha\footnote{Electronic address: dna@tifr.res.in},
Pankaj S. Joshi\footnote{Electronic address: psj@tifr.res.in}}

\affiliation{Tata Institute of Fundamental Research\\
Homi Bhabha Road, Mumbai 400005, India}

\begin{abstract}

In this paper we study gravitational lensing in the strong field limit from the perspective of 
cosmic censorship, to investigate whether or not naked singularities, if at all they exist in 
nature, can be distinguished from black holes. 
 The spacetime which we explore from this perspective is JMN metric which represents a
spherically symmetric solution to the Einstein field equations with anisotropic pressure and contains
a naked singularity at the center. JMN geometry is matched with the
\sch metric to the exterior at a finite radius.
 This metric was recently shown to be a possible end state of gravitational collapse of a 
fluid with zero radial pressure.
In the presence of the photon sphere gravitational
lensing signature of this spacetime is identical to that of \sch black hole with infinitely 
many relativistic images and Einstein rings, all of them located beyond a certain critical angle
from optic axis and the inner relativistic images all clumped together.
However, in the absence of the photon sphere, which is the case for a wide range 
of parameter values in this spacetime,  
we show that we get finitely many relativistic images and Einstein rings
spaced reasonably apart from one another, some of which can be formed inside the critical 
angle for the corresponding \sch black hole. This study suggests that the observation of 
relativistic images and rings might, in principle, allow us to unravel the existence
of the naked singularity in the absence of the photon sphere. 
 Also the results obtained here are in contrast with the earlier investigation on 
JNW naked singularities where it was shown that the radial caustic is always present in the absence
the photon sphere, which is not the case with JMN geometry where radial caustic is absent. 
We also point out the practical difficulties 
that might be encountered in the observation of the relativistic images and suggest that new
dedicated experiments and techniques have be developed in future for this purpose.

\end{abstract}
\pacs{95.30.Sf, 04.20.Dw, 04.70.Bw, 98.62.Sb}
\maketitle

\section{Introduction}
Deflection of light by massive bodies and therefore the phenomenon of gravitational 
lensing is a prediction of \gr which has helped in observationally testing  Einstein
theory against Newtonian gravity. In fact it was one of the first successfully 
verified predictions of \gr in 1919. Since then there has been numerous studies of 
gravitational lensing, both theoretically and observationally. But for good reasons, mostly these 
have been confined to the weak field approximation. The last decade, however, 
has seen a great rise in the interest in \grle in the strong field regime. This is 
important as a test of \gr in itself (because almost all generalizations of \gr should 
by construction reproduce the same weak field limit), as well as for the testing of 
various compact object scenarios in \gr (because they too have the same weak field).
 In particular, it is important to study whether and how strong 
field lensing can distinguish between various end states of gravitational collapse
of massive matter clouds (such as a massive star continually collapsing at the 
end of its life cycle), viz. black hole and naked singularity. This is also important
from the perspective of the cosmic censorship conjecture.

Cosmic censorship conjecture was proposed by Penrose in order to get rid of the naked singularities
in the real world around us \cite{pen69}. However the cosmic censorship conjecture is not yet proved
even several decades after it was put forward. There were many studies
recently where it was shown that the black holes as well as naked singularities are formed in a 
continual gravitational collapse of a matter cloud of reasonable matter field starting 
from a regular initial data \cite{Joshi1,Joshi2}. Thus naked singularities might occur in nature.
Their occurrence or otherwise is hard to infer from  purely theoretical investigations as it 
is extremely difficult to solve the Einstein equations in an astrophysically realistic scenario.
Thus one could take a phenomenological approach, where the consequences of the occurrence 
of the naked singularities computed theoretically are compared with the observations to either 
confirm or rule out the existence of the naked singularities.
In this paper we explore the gravitational lensing from such a perspective.
We note here that the strong gravitational lensing in JNW spacetime \cite{vnc98,nsl.ve}, lensing
in post-Newtonian framework for Kerr geometry \cite{wp} as well as for the rotating 
generalization of JNW spacetime \cite{gy} and the investigation of the shape and the position 
of the shadow in Kerr and Tamimatsu-Sato 
spacetimes \cite{bambi1,bambi2,maeda} has been done recently to address the same question.

Early works on strong field lensing were by Darwin 
\cite{dar59,dar61}, who studied the behavior of null geodesics 
in  strong field regime of \sch black hole and pointed out the divergence of \eda as 
the distance of closest approach of the geodesics approaches photon sphere.
Strong field lensing with a lens equation was studied by 
Virbhadra and Ellis \cite{sbhl.ve}, who examined strong field lensing in Schwarzschild
black holes and showed that there could in principle be infinite relativistic images
on each side of the black hole when a light ray with small enough impact parameter 
( distance of closest approach close enough to photon sphere) goes around  one or several times around 
the black hole before reaching the observer.
Earlier, lens equation for spherically symmetric static spacetimes that goes beyond
the weak field small-angle approximation was studied by Virbhadra ,Narasimha and Chitre in
 \cite{vnc98}.
The Virbhadra-Ellis type lens equation has also been applied to
boson star by D\c{a}browski and Schunck \cite{das00},
to a fermion star by Bili{\'c}, Nikoli{\'c} and Viollier
\cite{bnv00}. As one of the first steps towards using strong field 
lensing to probe the cosmic censorship question, Virbhadra and  Ellis have used this 
lens equation to study and compare  gravitational lensing by normal black holes and by naked
singularities modeled by the Janis, Newman, Winicour metric (JNW solution)\ct{nsl.ve}. 

It is worthwhile to extend this line of work to other novel, more interesting 
and if possible more realistic naked singularity models. With this in mind 
we consider here the class of solutions recently obtained by Joshi, Malafarina and Narayan 
\ct{jmn} as end state of certain dynamical collapse scenarios in a toy example. 
 JMN metric is a solution of Einstein field equations with an anisotropic pressure fluid and 
has a naked singularity at the center.
It is matched to the \sch metric at a certain radius. We refer to it here as JMN naked singularity from now on.
It is worthwhile to mention that, not only the presence of the central naked singularity but also the
value of the radius at which the interior solution is matched to exterior \sch geometry plays a crucial
role in determining gravitational lensing observables.

We should also mention that exact lens equations were proposed by \cite{fkn} for arbitrary 
spacetime and also by \ct{per04} for spherically symmetric case.
Bozza et al. have defined and analytically calculated strong field limit observables in
spherically symmetric spacetimes endowed with a photon sphere \cite{boz02,bcis01}.
In such a situation strong lensing from various alternatives/modifications of \sch geometry in
modeling the galactic center has been studied. For example lensing from regular black
holes was studied in \cite{eis11} and lensing from stringy black holes was studied in
\cite{bha03}. However the basic qualitative features in a lensing scenario in the presence
of a photon sphere is very similar to \sch case and is ineffective in probing the geometry 
beyond the photon sphere. Strong field lensing would be much easily able to probe 
differences from \sch spacetime if geometry being studied will be without a photon sphere. 
As we will see for the family of solutions studied in this paper, when the geometry has a
photon sphere the lensing signatures are exactly identical to \sch black hole case while
in the absence of photon sphere it is greatly different. 
In this work, the galactic supermassive compact object is analyzed as a 
strong gravity lens to illustrate these characteristics.

This paper is organized as follows. In section \ref{bf} we introduce the basic 
formalism in brief. In section \ref{galmod} we discuss the lens model with galactic 
supermassive dark object as the lens and in section \ref{sch} 
we discuss the lensing signatures when it is modeled as a \sch black hole. 
We discuss the naked singularity spacetime we intend to study and lensing 
in this background in \ref{jnm}  and compare this with \sch back hole and 
JNW solution in \ref{schcomp} \& \ref{jnwcomp} respectively. 
In section \ref{obs}, we discuss the implications of going beyond point source 
approximation for our study and in \ref{binary} we briefly discuss how binary 
systems could be useful for probing  question of cosmic censorship via gravitational 
lensing. Finally, we discuss the main results and conclude with a general discussion 
in section \ref{rem}.

\section{basic formalism}
\label{bf}
In this section we review the standard gravitational lensing formalism
\cite{vnc98,sbhl.ve} used in this paper to compute the location and
properties of the images.

We assume that the spacetime under 
consideration that is to be thought of as a gravitational lens is spherically symmetric, 
static and asymptotically flat. 
We assume that the source and the observer are located sufficiently far
away from the lens so that they can be taken to be at infinity for all practical purposes. 
We also assume that the source is a point-like object, 
although towards the end of the paper we describe how the results based on the point source assumption 
would change if the source has a finite extent instead of it being point-like.  
We assume that the geometrical optics approximation holds good. However we note that if we go arbitrarily 
close to the singularity, the Riemann curvature might become comparable to the 
wavelength of the light leading to the breakdown of the geometrical optics approximation.

The gravitational lensing calculations has two important parts. First one is the lens equation which relates
the location of the source to the location of the image given the amount of deflection suffered by the 
light from source to the observer as it passes by the the gravitational lens. The second important component 
is the deflection of the light encoded in the \eda $\hat\alpha(\theta)$ which we define later. 
We note that the deflection angle is the only input from the General theory of 
Relativity, and it can be computed by 
integrating the null geodesics.

\subsection{Lens equation}

\begin{figure}
\includegraphics[scale=.5]{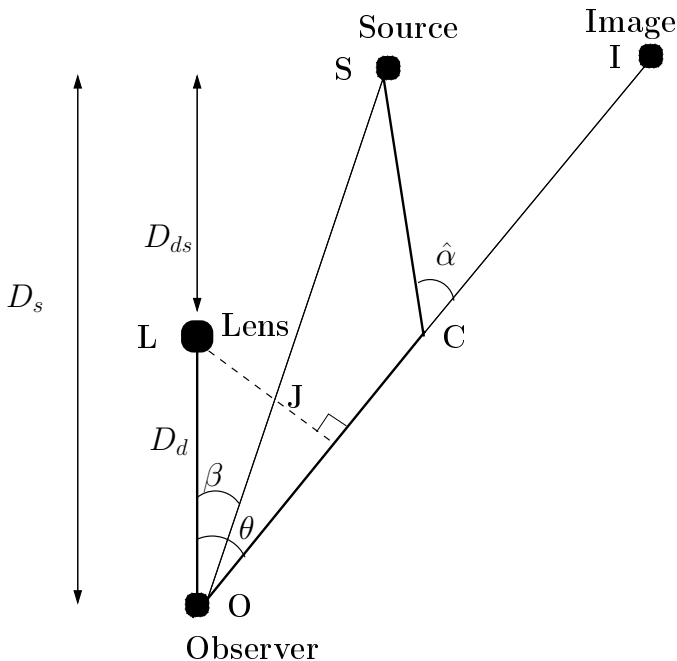} 
\caption{\label{lensd} The lens diagram: Positions of the source, lens, observer and the 
image are given by $S$, $L$, $O$ and $I$. The distances between lens-source, 
lens-observer and source-observer are given by $D_{ds}$, $D_{d}$ and $D_{s}$. 
The angular location of the source and the image with respect to optic axis are 
given by $\beta$ and $\theta$. The impact parameter is given by $J$. }
\end{figure}

The lens equation essentially  relates the position of the source to 
that of image. Fig\ref{lensd} is the lens diagram. It is same as lens 
diagram given in \cite{sbhl.ve}. 
The spherically symmetric spacetime under consideration is to be
thought of as a lens denoted by $L$ in the lens diagram.
The source $S$ and observer $O$ are located faraway as compared to the \sch radius, 
from the center of the spacetime in the asymptotic flat region. The line joining 
lens and the observer is known as the optic axis of the lensing geometry. In the absence of the 
lens light would have traveled along the line $SO$ and would have made an angle
 $\beta$ with respect to the optic axis. Thus $\beta$ is the source location. 
In the presence of the lens light gets bent. Let $SC$ and $OC$ be the tangents drawn 
to the trajectories of the light at the source and the observer. The angle $OC$ makes 
with the optic axis namely $\theta$ depicts the location of the image $I$. 
The angle $SCI$ is the Einstein deflection angle $\hat{\alpha}$ which we calculate later 
in this section. The distances from source to lens, observer to lens and source to observer
are given by $D_{ds}$,$D_{d}$ and $D_{s}$ respectively. 

The light mostly travels on the lines $SC$ and $CO$ except for the region close 
to lens where curvature is large and it suffers from a deflection. When the deflection is large, 
light can go around the lens multiple times.

Here we assume that $\beta$ is very small i.e. the observer, lens and the source are
aligned to a very good approximation. Let $LN$ be the perpendicular drawn to $OC$ from the lens.
$J$ here is the impact parameter. From the lens diagram we get 
\be
\sin\theta = \frac{J}{D_d}.
\label{geoeq}
\ee
The location of the source $\beta$ and the image $\theta$ can be related to each 
other by the following relation from the lens diagram.
\be
\tan\beta =  \tan\theta - \alpha ,
\label{LensEqn}
\ee
where 
\be
\alpha \equiv
    \frac{D_{ds}}{D_s}  \lt[\tan\theta + \tan\lt(\hat{\alpha} - \theta\rt)\rt].
\label{Alpha}
\ee
From the diagram above it it clear that what enters into the lens equation is the 
deflection angle $\hat\alpha$ modulo $2\pi$. 

We note that the many versions of the lens equation have been used in the literature
depending on the need and convenience. The lens equation \ref{LensEqn} used 
in this paper was derived by Virbhadra and Ellis \cite{sbhl.ve}. 
It allows for a arbitrarily large deflection of the light.
We note that one of the coauthors of this paper (DN) along with Virbhadra and Chitre \cite{vnc98} had worked on 
a different lens equation in a first investigation of the strong lensing 
phenomenon with large deflection;
but none of the features we describe here change if we use that equation.

\subsection{Deflection angle}
One requires the knowledge of the metric of the spacetime to derive the expression for
the Einstein deflection angle.
Consider general spherically symmetric static spacetime. The metric in the Schwarzschild-like
coordinates $\left(t,r,\nu,\phi\right)$ can be written as
\be
ds^2=- g(r)dt^2+\frac{1}{f(r)} dr^2
      +r^2d\Omega^2,
\label{Metric}
\ee
where $g(r)$ and $f(r)$ are arbitrary functions. Asymptotic flatness demands that 
$g(r\rightarrow \infty)=f(r\rightarrow \infty)=1$.

In a gravitational lensing scenario under 
consideration, source, observer and the lens define a plane.
In a spherically symmetric spacetime, the trajectory of the photon is confined to a plane
passing through the center which by the appropriate gauge choice can be taken to be 
the equatorial plane ($\vartheta=\pi/2$). Thus only those light rays 
emitted by the source, which travel in this plane can possibly reach the observer, 
ultimately leading to the formation of images and Einstein rings, and
this plane can be taken to be the equatorial plane without  loss of
generality.

The equation of motion for the light ray can be written as 
\be
U^t=\frac{1}{Jg(r)}, \ U^{\vartheta}=0, \ U^{\phi}=\frac{1}{r^2}
\label{Utpn}
\ee
and the radial motion is described by the equation
\be
\fr{g(r)}{f(r)}\lt(\fr{dr}{d \lambda}\rt)^2+V_{eff}(r)=\frac{1}{J^2}
\label{Ur}
\ee
where 
\be
V_{eff}=\fr{g(r)}{r^2}
\ee
can be thought of as an effective potential for the radial motion.
Here $U=\left(U^t,U^r,U^{\vartheta},U^{\phi}\right)$ stands for the velocity of the photon, 
$\lambda$ is the affine parameter and 
as stated earlier, $J$ is the impact parameter.

We can relate the impact parameter $J$ and the distance of closest approach $r_{0}$ 
using \eq{Utpn},\eq{Ur} and by setting $\frac{dr}{d\phi}=0$ in the following way
\be
J(r_{0}) = r_o  \sqrt{\frac{1}{g(r_o)}} .
\label{impp}
\ee
The total deflection suffered by the light ray as it travels from the source to the observer (i.e. 
the deflection angle) as a function of a distance of the closest approach of the 
light ray to the lens, is $\hat{\alpha}(r_{0})=2\int_{r_{0}}^{\infty}\frac{U^{r}}{U^{\phi}}d\phi-\pi$ 
So it is given by
\be
\hat{\alpha}\lt(r_0\rt)
 = 2 {\int_{r_0}}^{\infty}
\lt(\frac{1}{f(r)}\rt)^{1/2}
\lt[
\lt(\frac{r}{r_0}\rt)^2
 \frac{g(r_0)}{g(r)} -1
\rt]^{-1/2} \ \frac{dr}{r}  - \pi ,
\label{defl}
\ee

One important question for lensing in strong field regime is the presence/absence of photon
 sphere which is a $r=const$ null geodesics. As the distance of closest approach 
asymptotically approaches the photon 
sphere, the photon  revolves around the lens more and more number of  times and 
the bending angle $\hat\alpha$ diverges as the distance of closest approach tends 
to photon sphere $r_ph$. The maxima/minima of $V_{eff}$ give unstable/stable photon 
spheres. Thus equation for photon sphere is given by 
\be
\fr{d g(r)}{d r}=\fr{2 g(r)}{r}
\label{phh}
\ee

The effective potential  for photons in this geometry  gives an idea
of the radial behavior of photon trajectories, in particular the
turning points for photons. Hence it gives an idea as to when photons
coming from infinity get captured and when they can escape back to
asymptotia.
If $J>J(r_{ph})$ then the photon turns back from radius $r>r_{ph}$ before it reaches the photon sphere.
On the other hand if $J<J(r_{ph})$, the photon enters the photon sphere. 
Whether or not photon is captured or it again comes out after it has entered the photon sphere depends 
on the behavior of the effective potential in the interior of the photon sphere. 
Typically in most of the situations photon never comes out once it enters the photon sphere.



\subsection{Lensing observables}
We now describe the important lensing observables.
For a fixed position of the source, we compute the position of images and their magnifications.

All those values of $\theta$ that satisfy the lens equation \eq{LensEqn} for fixed 
values of the source position $\beta$ yield us the location of the images. 
In order to do that we must write down the deflection angle as
a function of the source position $\hat{\alpha}(\theta)$. 
This can be achieved using \eq{geoeq},\eq{impp}.

The cross-section of the bundle of rays gets modified due to the lensing. 
Liouville's theorem implies that the surface brightness is preserved. 
Thus the magnification i.e. ratio of the flux of the image to the flux of the 
source is the ratio of the solid angle subtended by the image to that 
of the source at the location of the observer.

The total magnification is defined as 
\be
\mu \equiv \lt( \frac{\sin{\beta}}{\sin{\theta}} \ \frac{d\beta}{d\theta} \rt)^{-1}.
\label{Mu}
\ee
which can be broken down into the tangential and radial magnification in the following way.
\be
\mu_t \equiv \lt(\frac{\sin{\beta}}{\sin{\theta}}\rt)^{-1}, ~ ~ ~
\mu_r \equiv \lt(\frac{d\beta}{d\theta}\rt)^{-1}
\label{MutMur}
\ee

The sign of the magnification of an image gives the parity of the image. 
The singularities of the tangential and radial magnification yield the tangential critical curves (TCCs)
and radial critical curves (RCCs), respectively in the lens plane and tangential caustic (TC)
and radial caustics (RCs) respectively in the source plane.

It is obvious from the expression for the tangential magnification that $\beta = 0$
gives the TC and the corresponding values of $\theta$ are the TCCs, also known 
as Einstein rings (ER). Thus Einstein rings can be obtained by solving for lens 
equation for $\beta=0$ i.e, in aligned configuration of source, lens and observer.

Using the lens equation \eq{LensEqn} the radial magnification \eq{MutMur} 
can be written in the following way:
\begin{eqnarray}
\frac{d\beta}{d\theta}=\left(1-\frac{D_{ds}}{D_{s}}\right)\frac{\sec^2{\theta}}{\sec^2{\beta}}-
\frac{D_{ds}}{D_{s}}\frac{\sec^2{(\hat{\alpha}}-\theta)}{\sec^2{\beta}}
\left(\frac{d\hat{\alpha}}{d\theta}-1\right)
\end{eqnarray}
It is clear from the expression above that if $\frac{d\hat{\alpha}}{d\theta}<0$ i.e. when $\hat{\alpha}$ is
a monotonically decreasing function, we have $\frac{d\beta}{d\theta}>0$ 
and the radial magnification will never diverge. Thus the radial critical curves would be absent, 
which will be the case for \sch as well as JMN naked singularity geometry dealt in this paper later .

\section{Galactic central supermassive object as a lens }
\label{galmod}
In this section we describe the gravitational lensing scenario that we 
would like to focus on in this paper.  The central supermassive dark object 
in our galaxy is modeled initially as \sch black hole and in the later 
section as a naked singularity. The mass of this object is taken to 
be $M  = 2.8 \times 10^6 M_{\odot}$ which is 
the mass of the supermassive black hole in our galaxy.
Distance of the source from the center of the galaxy is taken to be the distance of the sun from 
the galactic center $D_d =8.5 kpc$. Thus, in our example,
in the near-aligned configuration the lens is situated midway between 
the source and the observer i.e. $D_{ds}/D_s = 1/2$. 
Ratio of mass of the lens to the distance to the observer which would 
later appear in the calculations is $M/D_d \approx 1.57 \times 10^{-11}$.

\section{Gravitational lensing by \sch black hole}
\label{sch}
In this section we provide a brief overview of the results related to the gravitational lensing 
of light in \sch black hole geometry. As we discuss later, the naked singularity geometry under 
investigation in this paper matches  with the \sch geometry at the finite radial coordinate.
Therefore for the light rays that stay in the \sch regime all the time, the gravitational 
lensing is identical to that in \sch spacetime. In the next section we discuss change
in the gravitational lensing properties due to the presence of the naked singularity and make
a critical comparison with \sch results. The gravitational lensing by the \sch black hole was 
explored in detail in \cite{sbhl.ve}. We discuss the relevant details and results here.

The \sch metric is given by 
\be
ds^2=- \left(1-\frac{2M}{r}\right)dt^2+\left(1-\frac{2M}{r}\right)^{-1}dr^2
      +r^2d\Omega^2\ee
For convenience we work in the dimensionless variable $x=\frac{r}{2M}$. 
The distance of closest approach is $x_{0}=\frac{r_{0}}{2M}$.

\subsection{Deflection angle and photon sphere}
\begin{figure}[b]
\includegraphics[scale=.7]{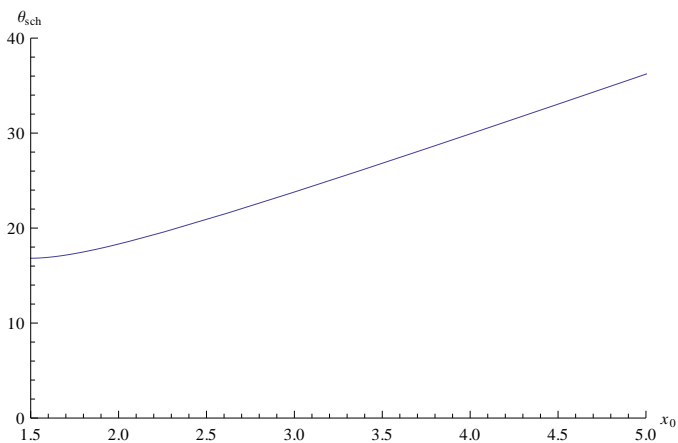}
\caption{\label{thxos} $\theta$ vs $x_0$ in \sch spacetime.
The image location $\theta$ (in \mas) is plotted as a function of closest approach $x_0$ (dimensionless) in \sch 
spacetime for a galactic supermassive black hole scenario. 
 It is a monotonically increasing function.
}
\end{figure}

We now compute the deflection angle as a function of the distance of minimum approach. From \eq{defl},
the Einstein deflection angle in the \sch spacetime in terms of the dimensionless variables is given by 
\begin{widetext}
\be
\hat{\alpha}\lt(x_0\rt)
 = 2 {\int_{x_0}}^{\infty}
\lt(\frac{1}{1-\frac{1}{x}}\rt)^{1/2} 
 \lt[
\lt(\frac{x}{x_0}\rt)^2
 \lt(\fr{{1-\frac{1}{x_0}}}{{1-\frac{1}{x}}}\rt) -1
\rt]^{-1/2} \frac{dx}{x}
-\pi
\label{defls}
\ee
\end{widetext}
As stated earlier the Einstein deflection angle diverges when the distance of minimum approach 
is very close to the radius of the photon sphere as the light circles around the center multiple times.
It turns out that there is a photon sphere in a \sch spacetime that can be obtained by solving \eq{phh}
which is located at the radius $r=r_{ph}=3M$ or in terms of dimensionless variable at $x=x_{ph}=1.5$.
Using \eq{geoeq} and \eq{impp} the distance of minimum approach $x_{0}$ can be translated 
into the image location $\theta$ as
\be
\sin\theta = \frac{2M}{D_d}\frac{x_0}{\sqrt{1-\fr{1}{x_0}}}
\label{thx0s}
\ee
which allows us to write deflection angle $\hat{\alpha}$ as 
a function of image location $\theta)$.
We have plotted $\theta(x_0)$ in Fig\ref{thxos}. It is a monotonically increasing function
of $x_0$.

\begin{figure}
\includegraphics[scale=.7]{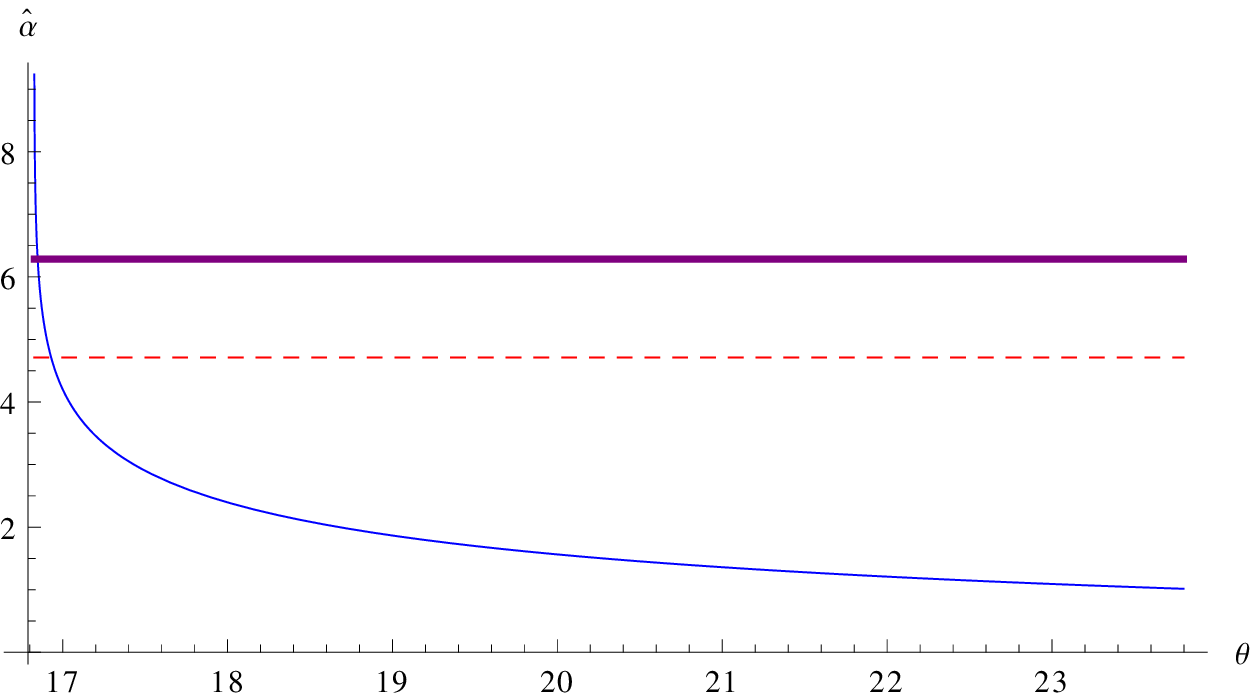}
\caption{\label{reld} $\hat{\alpha}$ vs $\theta$: Einstein deflection angle $\hat{\alpha}$ (in radian) 
is plotted as a function of the image location 
$\theta$ ( in \mas) in \sch spacetime for a Galactic supermassive black hole scenario. 
$\hat{\alpha}(\theta)$ is monotonically decreasing.
It diverges around $\theta\approx16.8$ \mas which corresponds to the photon sphere.
The horizontal lines correspond to $3 \pi /2$   and $2 \pi$ for
dashed(red) and thick(purple) lines respectively which mark the onset
of the relativistic deflection of light. 
Relativistic images and Einstein rings can be seen between the photon sphere 
and the intersection points of horizontal lines with $\hat{\alpha}$ curve. }
\end{figure}

We plot the Einstein deflection angle $\hat{\alpha}(\theta)$ in Fig\ref{reld}

We have made use here of the values of the different quantities we have chosen 
in a galactic central supermassive black hole scenario we discussed earlier.
It is a monotonically decreasing function of $\theta$. 
It diverges as we approach the photon sphere which corresponds to 
$\theta \approx 16.8$ microarcseconds.

\subsection{Images}
We now qualitatively describe the images formed due to the deflection of the light 
by \sch black hole in the galactic central supermassive object scenario. 
The images' locations can be obtained by solving the 
lens equation for the chosen source location. The image is said to be relativistic if 
the deflection of the light ray is larger than $3\frac{\pi}{2}$ as per the convention used in \cite{sbhl.ve}. 

In the weak field limit when the impact parameter is large and the deflection angle is small a pair
of nonrelativistic images are formed. They have opposite parities.
For small enough impact parameter, we get relativistic images with large deflection.
Theoretically there are infinitely many images formed on both sides of the optic axis i.e. 
with both positive and negative values of $\theta$. 
The relativistic images are bunched together around $\theta \approx 16.8$
\mas. This is an extremely important point that no images are formed between the the optic axis and 
$\theta \approx 16.8$ \mas. As we discuss later the situation can be significantly 
different in the case of the naked singularity.
More details on the location of the images and the magnification can be found in \cite{sbhl.ve}.

\subsection{Einstein rings}
As the Einstein deflection angle is a monotonically decreasing function as seen from Fig\ref{reld}, 
there is no radial critical curve present in the geometry. Location of the Einstein rings can be obtained 
by solving the lens equation with $\beta=0$. The Einstein rings are said to be relativistic if the 
deflection angle larger that $2\pi$ as per the convention used in \cite{sbhl.ve}. 

With our choice of parameters for galactic supermassive object scenario we get an 
equation $\tan{\theta}=\tan{(\hat{\alpha}-\theta)}$, which 
admits a solution $\hat{\alpha}=2n\pi+2\theta$.
There is a nonrelativistic Einstein ring which can be obtained by 
solving this equation for $\theta$ with $n=0$.
It is located at $\theta=1.15$ arcsecond.
There are infinitely many Einstein rings that can be obtained by solving the above 
equation with different values of $n$. All the relativistic Einstein rings are located close to
$\theta\approx 16.8$ \mas. As in the case of images there are no Einstien rings 
located between the optic axis and $\theta=16.8$ \mas. In case of the naked singularity geometry that 
we are about to discuss, the situation is significantly different.

\section{ Gravitational lensing by JMN Naked Singularity }
\label{jnm}
 In this section we study the gravitational lensing by a spacetime containing naked singularity.
We imagine a hypothetical situation where the galactic supermassive object
is modeled by a naked singularity 
solution described by a  specific metric. We study the images and the Einstein rings 
in the same situation and make a comparison. The spacetime geometry (to be 
referred to as JMN solution henceforth in this paper) we will be dealing with 
is a naked singularity solution obtained in 
\cite{jmn} as the end state of dynamical collapse from regular
initial conditions for a fluid with zero radial pressure but non-vanishing tangential pressure.
The solution  has a naked singularity at the center and matches  to a Schwarzschild spacetime
across the boundary $r=R_b$. Basic features of accretion disks in such a model was 
studied by the same authors and differences with black hole case were pointed out. 
In the same spirit, \grle in this background would also be an interesting observational 
probe of the toy model.

\subsection{JMN naked singularity geometry}
The spacetime is divided into two parts. 
(a) The interior region which is described by the following metric:
\begin{equation}
ds^2_e =  -(1-M_0)\left(\frac{r}{R_b}\right)^{\frac{M_0}{1-M_0}}dt^2+
\frac{dr^2}{1-M_0}+r^2d\Omega^2 \; .
\end{equation}
It can be easily shown that the curvature blows up at the center and 
thus it corresponds to a strong curvature time--like singularity.
(b) The exterior region is described by a Schwarzschild solution 
\begin{equation}
ds^2= -\left(1- \frac{M_0 R_b}{R}\right)dt^2+\frac{dr^2}
{(1- M_0 R_b/r)}+r^2d\Omega^2 \; .
\end{equation}
There is no event horizon in this geometry and thus the singularity at the center is exposed to the
asymptotic observer at infinity. Therefore it is a naked singularity.

It can be easily verified that the two metrics are connected across the boundary $R=R_b$ via $C^2$ matching.
There are two parameters in the solution, $M_0$ which is a dimensionless 
parameter and, $R_b$ which is the boundary radius at which the interior 
naked singularity metric is matched to the exterior \sch metric.
We must have $0<M_0<1$.
The \sch mass is related to these two parameters by a relation $M=\frac{M_0 R_b}{2}$. 
We fix the \sch mass to be same as we had chosen in the previous section for the sake of comparison.
Thus for fixed $M$ the only free parameter happens be $M_0$. The boundary between the two region is related 
to $M_0$ as $R_b=\frac{2M}{M_0}$.

As in the \sch black hole case we introduce the dimensionless variable $x=\frac{r}{2 M}$, 
where all the distances are expressed in units of the \sch radius. Thus the impact parameter
is $x_0=\frac{r_0}{2 M}$ and the boundary radius becomes simply the inverse of the parameter $M_0$ i.e.
$x_b=\frac{R_b}{2 M}=\frac{1}{M_0}$.
Thus larger the parameter $M_0$, smaller is the radius of the boundary for a 
given mass $M$ and we have a more compact object.

\subsection{Deflection angle}
We now compute the deflection angle. The first question one would like to
ask for lensing in strong field regime is whether photon sphere is present in the spacetime, since the 
bending angle $\hat\alpha$ would diverge as the distance of 
closest approach tends to $x_ph$ where $x=x_ph$ is the location of the photon sphere.

In order to investigate the existence of the photon sphere or otherwise and its location 
we look for the maximum of the effective potential $V_{eff}=\fr{g(r)}{r^2}$ or in other words
we solve the equation \eq{phh}.

\begin{figure}
\includegraphics[scale=.75]{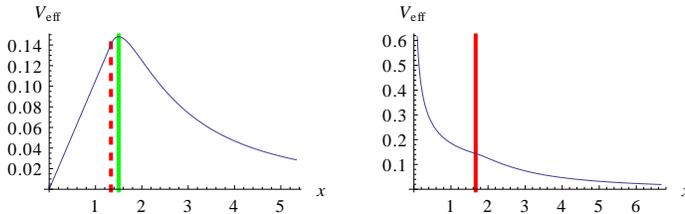} 
\caption{\label{veff} $V_{eff}$ vs $x$: Plotted here is the effective potential $V_{eff}$ (dimensionless) 
for the radial motion for the light rays as a function of $x$ (dimensionless). 
The plot on the left corresponds to the parameter values $M_0 \ge 2/3$. The thick(green) vertical
line corresponds to the photon sphere, whereas the dashed(red) line corresponds to the boundary between the interior
naked singularity region and exterior \sch geometry. The effective potential monotonically 
goes to zero in the interior region. 
The plot on right corresponds to the parameter values $M_0 <2/3$, where the photon sphere is absent.
The vertical red line is the boundary. The effective potential blows up as we approach the singularity.}
\end{figure}


When $M_0 \ge \frac{2}{3}$ the boundary between the interior and exterior \sch region is at 
$x_b=\frac{1}{M_0}\le \frac{3}{2}$. So the boundary is below the photon sphere in the exterior 
\sch region $x \le x_{ph}$. The effective potential for the radial motion of the light ray is 
plotted in Fig\ref{veff}(left part). The effective potential goes on decreasing below the photon sphere
and asymptotically goes to zero. It is clear from the behavior of the effective potential that 
the light ray which enters the photon sphere never turns back and it eventually hits the naked singularity.
Thus it is captured. Photons can turn back from the region exterior to the photon sphere and 
deflection angle goes on increasing indefinitely as we approach approach it. 
This implies that the lensing will be exactly identical to \sch case  
which we discussed in the previous section
as it is not possible for photons coming from and going back to a large distance probe the 
metric interior to photon sphere. Thus the gravitational lensing 
cannot unravel the possible existence of the naked 
singularity at the center for $M_0 \ge \frac{2}{3}$.

When $M_0 < \frac{2}{3}$ the boundary is located at $x_b=\frac{1}{M_0}> \frac{3}{2}$. Thus there
is no \sch photon sphere in the exterior region, since the boundary is outside the location of the \sch
photon sphere $x_b>x_{ph}$.
The effective potential for the radial motion for the light rays is plotted in Fig\ref{veff}(right part).
The effective potential is monotonically decreasing function in the interior region. 
Since it does not admit any extremum, no photon sphere is present in the interior
 region as well. The effective potential 
in fact blows up at the singularity which implies that no light ray can reach it. 

From now on wards we focus on the case $M_0 < \frac{2}{3}$. If the distance 
of minimum approach is larger than $x_0 \ge x_b=\frac{1}{M_0}$, 
then the light ray travels in the exterior \sch geometry and the images and Einstein
rings formed due to the lensing are identical to those discussed in the previous section. 
Thus we focus on the case where the distance of minimum approach is 
less that the boundary radius $x_0 < x_b=\frac{1}{M_0}$.
So that the light rays travel partly in the external \sch metric and partly in the interior metric 
and it can actually probe the interior, containing  naked singularity. We would like to understand the
formation of images and the Einstein rings due to the light rays passing through the interior region.

The Einstein deflection as a function of distance of closest approach when  $x_0<\frac{1}{M_0}$ is given by 
\begin{widetext}
\be
\hat{\alpha}\lt(x_0\rt)
 = 2 {\int_{x_0}}^{\frac{1}{M_0}}
\lt(\frac{1}{1-M_0}\rt)^{1/2}
\lt[
\lt(\frac{x}{x_0}\rt)^2
 \lt(\frac{x_0}{x}\rt)^{\gamma} -1
\rt]^{-1/2} \frac{dx}{x}+  2 \int_{\frac{1}{M_0}}^{\infty}
\lt(\frac{1}{1-\frac{1}{x}}\rt)^{1/2}
\lt[
\lt(\frac{x}{x_0}\rt)^2
 \frac{(1-M_0)(x_0 M_0)^{\gamma}}{1-\frac{1}{x}} -1
\rt]^{-1/2} \frac{dx}{x} - \pi
\label{deflj}
\ee
\end{widetext}
where
$\gamma =\frac{M_0}{1-M_0}$. The first term corresponds to the contribution to the deflection angle 
from the interior region and the second term corresponds to the contribution 
from the exterior \sch region. 

\begin{figure}[b]
\includegraphics[scale=.7]{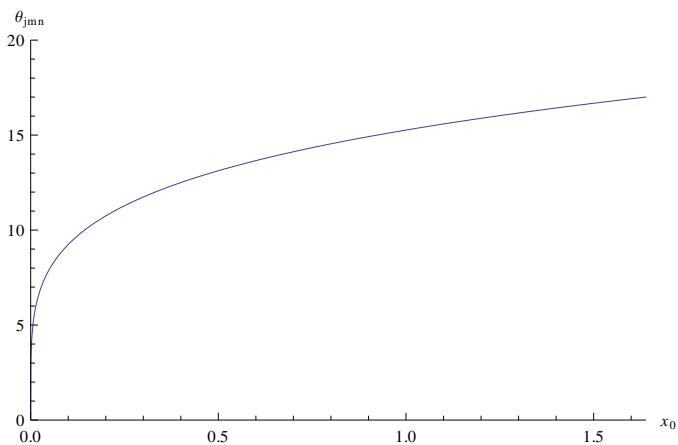}
\caption{\label{thxoj} $\theta$ vs $x_0$ in JMN: The image location $\theta$ (in \mas) is plotted as a function 
of closest approach $x_0$ (dimensionless) in JMN spacetime for a galactic central 
supermassive object scenario in an interior region. 
It is a monotonically increasing function.
}
\end{figure}

\begin{figure}
\includegraphics[scale=.7]{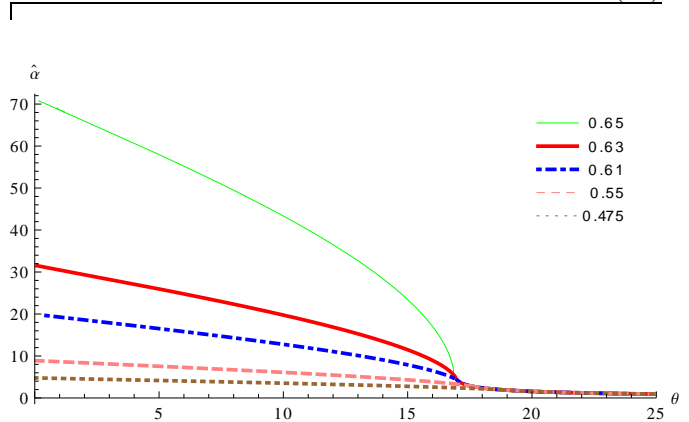}
\caption{\label{alphahat} $\hat{\alpha}$ vs $\theta$ for different $M_0$:
 Einstein deflection angle $\hat{\alpha}$ (in radian) in JMN spacetime is plotted 
against the image location $\theta$ (in \mas).
The value of $M_0$ for the curves are shown in the legends.
$\hat{\alpha}$ is a monotonically decreasing function. As we increase $M_0$ the maximum
deflection angle goes on increasing. The deflection angle will be larger than $2\pi$ and thus relativistic
images can form for $M_0>0.475$. The curves for different values of $M_0$ cross because ultimately
all the curves have to match to the deflection angle curve for \sch spacetime beyond the boundary and since 
boundary $x_b=\frac{1}{M_0}$ is at the smaller radius for larger $M_0$.}
\end{figure}

Using the relationship between $\theta$ and $x_0$, i.e.
\be
\sin{\theta} = \frac{2M}{D_d}\frac{x_0}{\sqrt{(1-M_0)(x_0 M_0)^{\gamma}}}
\label{thx0j}
\ee
we can express deflection angle as a function of image location $\hat{\alpha}(\theta)$.
We have plotted $\theta(x_0)$ for the geometry in Fig\ref{thxoj} and $\hat{\alpha}(\theta)$  in Fig\ref{alphahat}.

Presence of the photon sphere guarantees the relativistic deflection of light,
though, as we pointed out earlier, it prevents probes to the interior density structure.
But we are investigating the 
parameter regime $M_0 < \frac{2}{3}$ where the photon sphere is absent in the geometry. 
So the relativistic deflection may or may not occur. Firstly we would like to find out 
the parameter range $M_0$
where we expect relativistic deflection of light to happen. 

Looking at the Fig\ref{reld} and Fig \ref{alphahat}, \eda is monotonically increasing
with decreasing $x_0$ for both Schwarzschild and JMN.
In Schwarzschild geometry the deflection angle reaches $\fr{3 \pi}{2}$ at
$x_0\sim 1.605$ i.e. if $M_0>0.62$. 
So any $x_b$ less than that (and hence $M_0 >0.62$) will definitely give 
relativistic defection. 
This is sufficient but not necessary condition. 
More detailed calculation shows that the maximum value of the deflection 
angle is larger than $2\pi$ for $M_0>0.475$. 
Thus relativistic images and Einstein rings can form beyond this parameter value. 
Note that this is independent any value of Schwarzschild mass.


\subsection{Images}
We now describe the properties of the images formed due to the gravitational 
lensing of the light passing through the interior region. 
The relativistic images are possible only beyond the parameter value $M_0>0.475$.
These images probe the interior geometry and can unravel existence of the naked singularity.

We calculate the location of the images and their magnification 
for galactic supermassive object scenario with a 
given source location that is in the near-aligned configuration. We solve the
lens equation for fixed $\beta=0.075 $ and for given $M_0>0.475$. 
For different values of $M_0$ we get different number of
images on the same side as well as on the opposite side of the optic axis. 
The number of image goes on increasing as we increase $M_0$. 

In Table \ref{same},\ref{opp} we make a list of image locations and magnifications for 
a specific value of $M_0=0.63$. 
There are four images on the same same side as well as on opposite side of the
optic axis. Images are well separated from one another with angular separation of
around 2-5 microacrsecond.
The magnification of all the images is of the same order of magnitude. 
The radial parity of the image is always
positive. The tangential parity and thus the total parity is positive for the images 
on the same side of the optic axis while it is negative for the images on the opposite side. 
Also it is worthwhile to mention here that the position of relativistic images 
does not change much with changes in source position which can also be inferred 
from the fact that the radial magnification of the images  (which is $\fr{d\th}{d\beta}$) 
is of the order of $10^{-12}$ as shown in tables \ref{same} and \ref{opp}.

\begin{table}

\caption{Images and magnification (same side): In this table we list the location 
of the relativistic images and magnifications 
on the same side as source for the galactic central supermassive object scenario
for $M_0=0.63$ and $\beta=0.075$ radian. $\th$ is in \mas. The images are well separated and 
have comparable magnifications}
    \begin{tabular}{|c|c|c|c|c|}
        \hline
        Image  & $\theta$ & $\mu_t$            & $\mu_r$            & $\mu$              \\ \hline \hline
        I      & $16.74$  & $1.9 \times 10^{-9}$ & $1.8\times 10^{-12}$ & $1.9\times 10^{-21}$ \\ 
        II     & $14.56$  & $0.9\times 10^{-9}$  & $4.7\times 10^{-12}$  & $4.4\times 10^{-21}$ \\ 
        III    & $10.75$  & $0.7\times 10^{-9}$  & $6.8\times 10^{-12}$ & $5.0\times 10^{-21}$ \\ 
        IV     & $5.82$   & $0.4\times 10^{-9}$  & $8.0\times 10^{-12}$ & $3.2\times 10^{-21}$ \\
        \hline
    \end{tabular}

\label{same}
\end{table}

\begin{table}

\caption{Images and magnification (opposite side): In this table we list the 
location of the relativistic images and magnifications 
on the opposite side as source for the galactic central supermassive object scenario 
for $M_0=0.63$ and $\beta=0.075$ radian. $\th$ is in \mas. The images are well separated and 
have comparable magnifications.}

    \begin{tabular}{|c|c|c|c|c|}
        \hline
        Image & $\theta$ & $\mu_t$               & $\mu_r$               & $\mu$                \\ \hline  \hline
        I     & $16.68$  & $-1.1 \times 10^{-9}$ & $1.9\times 10^{-12}$ & $-2.1\times 10^{-21}$ \\ 
        II    & $14.41$  & $-0.9\times 10^{-9}$  & $4.8\times 10^{-12}$ & $-4.4\times 10^{-21}$ \\ 
        III   & $10.54$  & $-0.7\times 10^{-9}$  & $6.9\times 10^{-12}$ & $-4.7\times 10^{-21}$ \\ 
        IV    & $5.58$   & $-0.4\times 10^{-9}$  & $8.1\times 10^{-12}$ & $-3.0\times 10^{-21}$ \\
        \hline
    \end{tabular}

\label{opp}
\end{table}

\begin{table}

\caption[]{Einstein rings: In this table we list the location of the relativistic Einstein rings for 
$M_0=0.63$ for the galactic central supermassive scenario and the corresponding
 values of the deflection angle, $\th$ is in \mas and $\hat{\alpha}$
 is in radian. The Einstein rings are well separated.}

    \begin{tabular}{|c|c|c|}
        \hline
        No. & $\theta_E$ & $\hat{\alpha}$ \\ \hline \hline
        I   & $16.715$   & $2\pi+0.00068$ \\ 
        II  & $14.485$   & $4\pi+0.00042$ \\ 
        III & $10.469$   & $6\pi+0.00076$ \\ 
        IV  & $5.700$    & $8\pi+0.00173$ \\
        \hline
    \end{tabular}

\label{er}
\end{table}

\subse{Einstein rings }
Since the deflection angle is a monotonically decreasing function of the image 
location the radial critical curves would be absent. The caustic happens to be a point $\beta=0$,
since we are dealing with spherically symmetric spacetimes. In order to compute the critical curves i.e.
the location of the Eisntein rings we solve the lens equation with $\beta=0$.
As discussed in the \sch case for the galactic supermassive object scenario,
 we have to solve the equation $\tan\th=\tan (\hat{\alpha}-\th)$, 
which holds good when $\hat{\alpha}=2n \pi+2 \th$. Knowing $\hat{\alpha}$ we can solve 
the previous equation to get the angular locations of the Einstein rings. 
For a given value of $M_0$, the number 
of solutions to this equation would be either 
 $\lt[\fr{\hat{\alpha}_{max}}{2\pi}\rt]$ or $\lt[\fr{\hat{\alpha}_{max}}{2\pi}\rt]-1$
which will be the number of the relativistic Einstein rings. 
The number of the relativistic Einstein rings goes on increasing as we increase $M_0$.
As in the \sch case there will be a nonrelativistic Einstein ring 
located at $1.15$ arcsecond which corresponds 
to the solution of the equation with $n=0$.

For $M_0=0.63$ we have four relativistic Einstein rings. The location of the relativistic Einstein rings 
and the corresponding Einstein deflection angles is given in Table
\ref{er}. The rings are well separated from one another with
separation of the order of 2-5 microarcseconds.

In Fig \ref{mut}, \ref{mur}, \ref{mu}, we show the variation of the tangential, radial and total 
magnification with $\theta $ near the relativistic Einstein rings.
As expected tangential and consequently total magnification diverges at the Einstein rings and 
falls rapidly as we move away from it. 

\begin{widetext}

\begin{figure}
\includegraphics{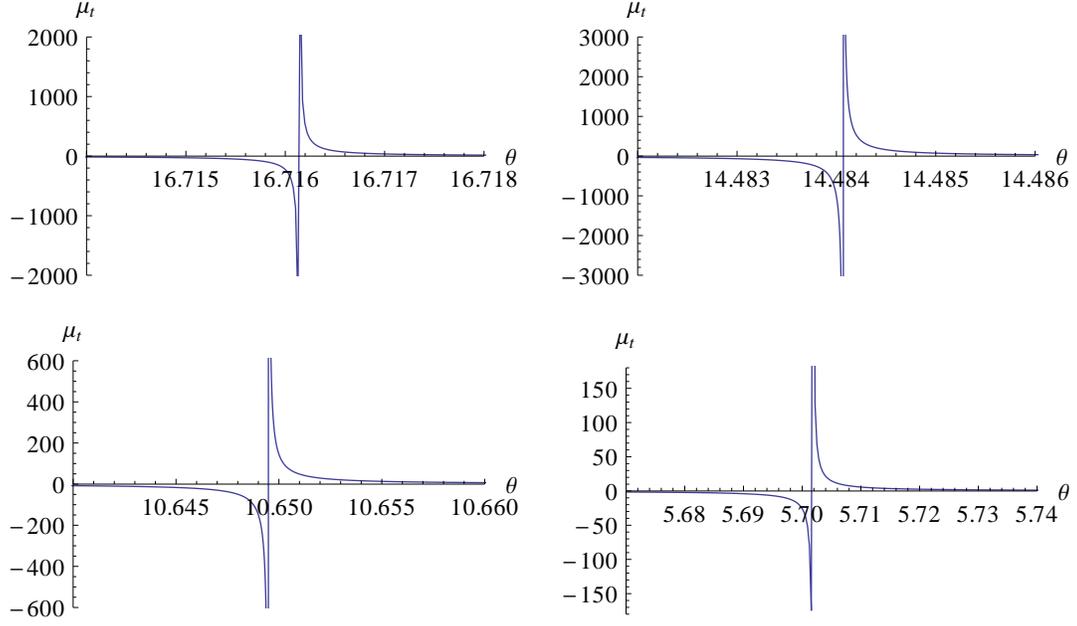}
\caption{\label{mut} $\mu_{t}$ vs $\theta$ for JMN spacetime:
Plotted here is the tangential magnification near the Relativistic 
Einstein Rings for $M_0=0.63$. $\th$ is in \mas and y-axis scale has 
been multiplied by $10^9$ for clarity. Tangential magnification blows up at the 
location of the Einstein ring and decreases as we go away from it.}
\end{figure}

\begin{figure}
\includegraphics{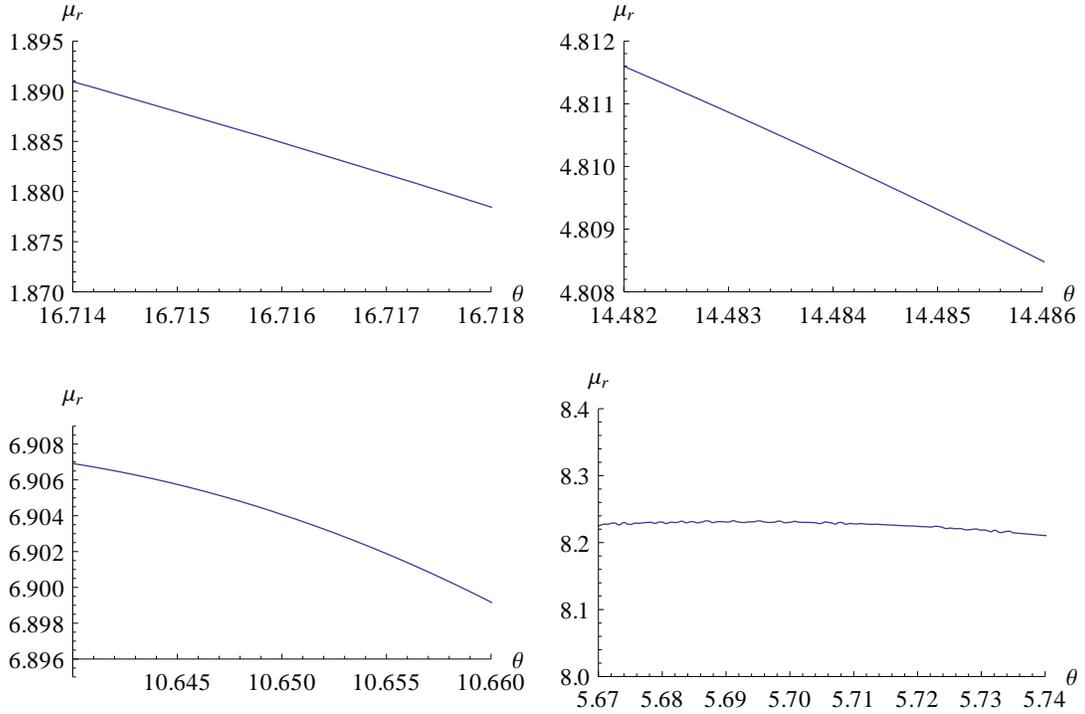}
\caption{\label{mur} $\mu_{r}$ vs $\theta$ for JMN spacetime:  Plotted 
here is the radial magnification near the Relativistic Einstein Rings for 
$M_0=0.63$. $\th$ is in \mas and y-axis scale has been multiplied by $10^{12}$ for clarity}
\end{figure}

\begin{figure}
\includegraphics{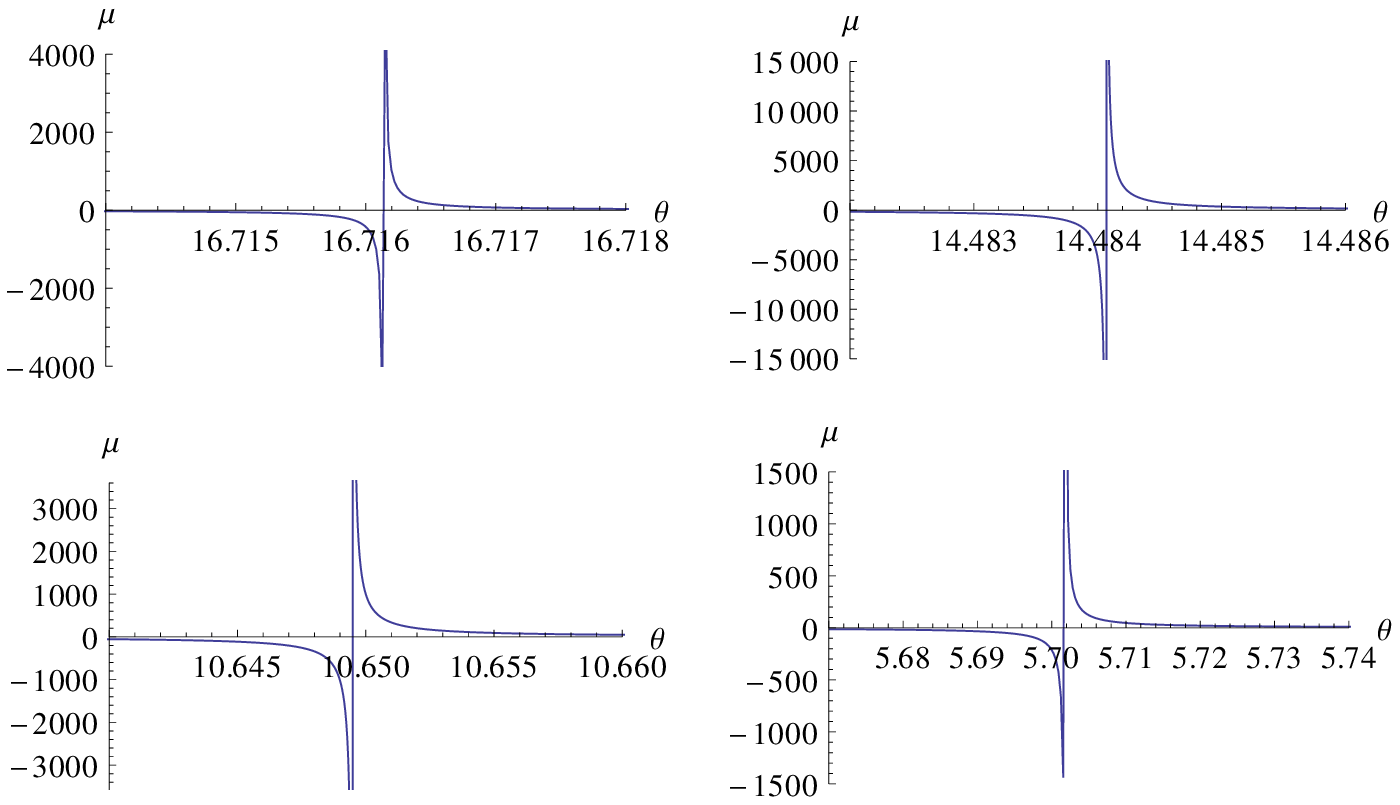}
\caption{\label{mu} $\mu$ vs $\theta$ for JMN spacetime: Plotted here
  is the total magnification near the Relativistic
 Einstein Rings for $M_0=0.63$. $\th$ is in \mas and y-axis scale has
 been multiplied by $10^{21}$ for clarity. Total magnification blows
 us at the
 location of the Einstein ring and decreases as we go away from it.}
\end{figure}

\end{widetext}


\section{Comparison with \sch black hole}
\label{schcomp}
In this section we make a critical comparison of the results obtained for gravitational lensing 
in \sch black hole and naked singularity geometry described by JMN solution with the same mass 
in a galactic central supermassive object scenario.

In the \sch black hole case the photon sphere is present. Apart from the outer nonrelativistic 
images and Einstein ring, infinite number of relativistic images and Einstein rings are formed. 
But they are clumped together around $\theta \approx 16.8$ \mas. No images or rings are formed between
the optic axis and  $\theta \approx 16.8$ \mas which happens to be the forbidden region. 
The separation between the
first and rest of the images clumped together is of the order of $0.1$ \mas 
and the ratio of magnifications is of the order of 500 \ct{sbhl.ve,bcis01,boz02}.

In JMN geometry when $M_0 \ge \frac{2}{3}$ \sch photon sphere is present and the gravitational lensing 
is identical to that of \sch black hole. If $M_0 \le 0.475$ no relativistic images are formed. 

The interesting regime in parameter space is $0.475<M_0<\frac{2}{3}$. A number of relativistic images 
and Einstein rings are formed depending on how large is $M_0$. Unlike \sch black hole case 
the images are not clumped together. They are well separated from one another. Images and Einstein rings 
can appear in the region between optic axis and $\theta \approx 16.8$ \mas which is a forbidden region 
for \sch black hole. The images have comparable magnifications. 

Thus there are qualitative differences in the images formed in \sch black hole geometry and 
JMN spacetime when $M_0<\frac{2}{3}$. 
These features can in principle lead to the observational distinction between the two spacetimes.

\section{comparison with JNW naked singularity}
\label{jnwcomp}
In the previous section we compared the gravitational lensing in \sch black hole geometry with that 
of JMN naked singularity. In this section we make a comparison between JMN naked singularity 
and JNW naked singularity which has been studied in the past from the perspective of strong gravitational
 lensing \cite{vnc98}
,\cite{nsl.ve}.
It is a solution of Einstein equations 
with a minimally coupled massless canonical scalar field with two parameters, mass $M$ and scalar charge $q$.
For low scalar charge $\frac{q}{M}\le \sqrt{3}$ this solution has a photon sphere. 
In this case the qualitative features of lensing are very similar to \sch case. 

However, for large 
scalar charge $\frac{q}{M}>\sqrt{3}$ photon sphere is absent in JNW spacetime. 
It was stated in \cite{nsl.ve} that the relativistic deflection and images are absent completely in the absence
of the photon sphere. However, a careful investigation in this range of parameters shows that in an extremely 
small range of parameter $\sqrt{3}<\frac{q}{M}\le 1.746$, relativistic
lensing and images are formed even in the absence
of the photon sphere. Relativistic images are absent when $\frac{q}{M}>1.746$. But interestingly
as mentioned in \cite{nsl.ve} the radial caustic is always present in the absence of the photon sphere.
This is a consequence of the fact that as we decrease the impact
parameter the deflection angle initially increases, it attains 
a maximum and goes on decreasing. Eventually it settles down to a constant negative value $-\pi$ in the limit where
impact parameter approaches zero.


 In the JMN spacetime, in the absence of the photon sphere we get
  relativistic images in a rather 
wide range of parameter values
$0.475<M_{0}<\frac{2}{3}$ as compared to JNW case. Relativistic images
are absent for $M_{0}<0.475$. However, 
radial caustic is always absent in JMN geometry. Thus the radial
caustic  will allow us to 
distinguish between JNW and JMN
naked singularities in the absence of the photon sphere.

We note that the massless scalar fields are not observed in nature. Also 
it is not clear whether or not JNW solution is an endstate of the gravitational collapse.
Thus it would be worthwhile to study other realistic solutions containing naked singularities
which might be an endstate of the gravitational collapse to see whether they exhibit lensing signature
which is quite different from JNW spacetime. 
The JMN spacetime that we study in this paper is a toy model with zero radial pressure, but it has the 
merit of being obtainable as a possible endstate
of the gravitational collapse. 
As we have shown, the gravitational lensing in 
JMN spacetime can be significantly
different from that in JNW geometry in the appropriate parameter range. 
In future we intend to study the lensing in the more realistic spacetime with nonvanishing radial pressure
and also which is the endstate of gravitational collapse.


\section{Observability of relativistic images in a realistic scenario}
\label{obs}
So far we studied the gravitational lensing in an idealized scenario and showed 
that JMN naked singularity can in principle be distinguished from \sch black hole 
and JNW naked singularity by the observation of the 
relativistic images and Einstein rings. In this section we try to gauge the possibility of observation 
of the relativistic images
in a realistic scenario with the current state of art instruments and techniques.

The distance of relativistic images from the optic axis is of the order of $ \frac{M}{D}\approx 10$ \mas and 
their magnification is  
of the order of $\mu \approx 10^{-21}$.
The angular distance from the optic axis is extremely small. Larger the mass and smaller the 
distance of the observer to the lens, larger will be the angular separation from the optic axis.
It might be possible to achieve \mas resolution
 with Very Long Base Interferometry
(VLBI) \cite{VLBI}. 
However since the magnification is extremely small the observation of these images
seems difficult. 

As we discussed earlier, magnification of the Einstein rings with point sources is infinite.
However in reality we deal with the sources of finite extent. 
We now estimate the expected values of the magnification 
for Einstein rings with source which is taken to be a star like the
 sun with radius $R \approx 7*10^8 m$. As 
discussed earlier in the galactic central supermassive object scenario,
 the distance of the sun as well as the sun-like star which acts as a source is 
$D\approx 8.5 kpc$. Orbital velocity of such a star will be comparable 
to that of sun  $v$ $\sim$ $220$ km/sec. 
The angle subtended by the star at the observer in a near aligned lensing situation 
would be $\Delta \beta=\frac{2R}{D}\approx 1$ \mas.
The angular radius of the
outer non-relativistic Einstein ring as in the case of \sch spacetime 
would be around $\theta_{E}\approx \sqrt{\frac{M}{D}} \approx 1 arcsec$,
whereas radius of the Einstein rings formed due to the relativistic deflection
would be $\theta_{E}\approx \frac{M}{D} \approx 10$ \mas.
In a realistic scenario a star with the finite extent 
is not stationary
but moves around with the orbital velocity and therefore the Einstein rings would appear 
only when the star crosses the caustic $\beta=0$ and would last for a 
time $\Delta T=\frac{R}{v}\approx 3200 sec\approx 50 min$,
whereas the time delay between the arrival of photons along the different trajectories
leading to different rings would be of the order 
of $\tau \approx \frac{GM}{c^3}\approx 10 sec$. 
Thus all the rings would appear and disappear almost simultaneously. 
The tangential magnification for a finite source is not infinite but takes 
an approximate value $\mu_{t}\approx \frac{\theta_{E}}{\Delta \beta}$. 
This expression can be derived easily for the external image by computing an averaged
tangential magnification for the finite size source. 
The average radial magnification is of the same order as the radial magnification 
at Einstein ring as it varies very slowly with image position.
Thus the tangential magnification of the outer non-relativistic Einstein ring would be
$\mu_{t}\approx 10^{6}$, whereas that of inner relativistic Einstein rings 
would be $\mu_{t}\approx 10$. 
The radial magnification of non-relativistic ring is of order $1$ while 
that of relativistic rings is $10^{-12}$. 
Then the total magnification of the outer rings is seventeen orders of magnitude 
larger than that of the inner rings. So it is doubtful that inner rings 
will be visible in the presence 
of the outer ring which is much more brighter with the current state of art instruments 
which may not have sensitivity over such a large range of magnification.

It seems that 
it may not be possible to unravel the true nature of the galactic central supermassive object 
at present as the inner images which carry the information about the 
possible existence of the naked singularity are swamped by the bright outer image which 
provides the information only about the Schwarzschild mass. We note that this is 
always going to be a generic feature associated with relativistic images and would be independent of the
specific naked singular metric under consideration. 

However keeping this point in mind dedicated instruments and techniques could be developed 
in the future which might be able to unravel the possible existence of a naked 
singularity instead of black hole.

\section{Gravitational lensing due to a binary system as a probe of cosmic censorship}
\label{binary}
In this section we propose a scenario where one might be able to test the cosmic censorship
with the current state of instruments and technology available to us. 
The assumption of the spherical symmetry made in this paper for the sake of simplicity is unrealistic. 
In the absence of spherical symmetry the caustic will be fairly complicated and not just a point. 
For instance one could imagine a situation where the evolution of the massive binary 
star system has eventually 
led to the configuration consisting of the naked singularity and
an ordinary star orbiting around each other. 
If one of the star 
is massive as compared to the other it will die much earlier. As an example 
50 $M_{\odot}$ star would die in $0.5$ million years and 
could possibly turn into a naked singularity instead of a black hole.
The main caustic in this scenario could be diamond shaped with cusps. A given source might cross 
caustics multiple times giving rise to the appearance and disappearance of
 the Einstein rings or large magnified images multiple times. 
Time difference between the disappearance and reappearance of the consecutive Einstein 
rings will be influenced by the presence of the naked singularity in the binary system. 
Thus it could be used to infer the presence of naked singularity, as against a black hole.  
Such a situation is however extremely difficult to model. It is beyond the scope of 
this paper and might be dealt with later.

\section{conclusion and discussion} 
\label{rem}
In this paper we studied the strong gravitational lensing from the perspective of cosmic censorship
and explored the possibility of distinguishing black holes from naked singularities.
We modeled the galactic central supermassive dark object initially by a 
black hole and then by naked singularity. We studied the gravitational lensing of the source 
in a near aligned configuration at a distance from a galactic center
approximately comparable to distance of the sun from the center.

The \sch black hole has a photon sphere. Thus apart from a pair of
nonrelativistic images and a nonrelativistic Einstein ring, infinitely many relativistic images
and Einstein rings clumped together. 
No images and Einstein rings lie in the region between the optic axis and 
$\theta=16.8$ \mas. Also all the images that are clumped together are 
highly demagnified as compared to the first 
relativistic image with a small separation between them of the order of $0.1$ \mas.

We then model the galactic center object as JMN solution which was recently shown to occur as an end state 
of the gravitational collapse of a fluid with zero radial pressure but non-vanishing tangential pressure.
This solution has two parameters, namely mass and another parameter $M_0$. The spacetime is divided
 into two parts.
Exterior metric is \sch spacetime with same mass as that of the \sch black hole considered earlier.
Interior metric contains a central naked singularity with the boundary located at the radius
$R_b=\frac{2M}{M_0}$.The two metrics are connected across the boundary by $C^2$ matching. 

In the parameter range $M_0 \ge \frac{2}{3}$ the \sch photon sphere is present 
in the geometry and the gravitatioanl lensing signature of JMN spacetime is identical to that of the 
\sch black hole.

When $M_0 \le 0.475$, the photon sphere is absent. But no relativitic bending of light and thus no
relativistic images possible. This behavior is different from the \sch black hole. 

The interesting parameter range is when $0.475<M_0 <\frac{2}{3}$. The photon sphere is absent. 
But the relativistic images and Einstein rings can form and their number increases with 
increasing value of the parameter $M_0$. The images and rings are well separated from 
one another and happen to lie in the forbidden region for \sch black hole, within a distance
from the optic axis 
of $\theta=16.8$ \mas. Their magnification is also comparable. Thus the strong
gravitational lensing signature is qualitatively different from \sch black hole. 


The gravitational lensing in the absence of the photon sphere is 
qualitatively different in JMN and JNW spacetimes.
In both the geometries relativistic images are present in an
appropriate parameter 
range. However, there are no radial caustics in
the JMN geometry, while radial caustic is always present in the 
JNW spacetime in the absence of the photon sphere.

However, there are practical difficulties as far as observation of relativistic images and rings are concerned
with the telescopes and techniques currently being used. 
We require \mas resolution which can be achieved with VLBI. However magnification of the 
images which is of the order of $\mu=10^{-22}$ is too small. Relativistic Einstein rings formed 
due to the lensing of the star with the size comparable to sun,
will be $10^{17}$ times weaker as compared to the nonrelativistic \sch Einstein 
ring and thus will not be 
seen since the current instruments do not have dynamical 
range over seventeen orders of magnitude of brightness.

Keeping this in mind, new techniques and instruments must be developed in the future which will be able to 
observe the Einstein rings and can unravel the nature of the galactic central supermassive object.

We also suggest that the appearance and disappearance of the outer Einstein ring as the source crosses
diamond shaped caustic more than once can possibly shed light on the possible existence of 
the naked singularity in the binary system of a naked singularity and a massive star. We wish to
explore this situation in the future.

In this paper we studied a naked singularity geometry arising out of a toy calculation of dynamical collapse
of a matter with only the tangential pressure. It would be interesting to study more realistic cases e.g.
with the inclusion of the radial pressure.

\section*{Acknowledgments}
We thank K.S.Virbhadra and the anonymous referee for valuable comments and suggestions.


\end{document}